\documentclass[journal=jpca,manuscript=article,layout=twocolumn]{achemso}

\usepackage[colorlinks,linkcolor=blue,citecolor=blue,urlcolor=black,bookmarks=false,hypertexnames=true]{hyperref} 
\usepackage{chemformula} % Formula subscripts using \ch{}
\usepackage[T1]{fontenc} % Use modern font encodings
\usepackage[version=3]{mhchem}
\usepackage{amssymb}
\usepackage{amsmath}
\usepackage{color}
\usepackage{braket}
\usepackage{xspace}
\usepackage{cleveref}
\usepackage{graphicx}
\usepackage{subfigure}
\usepackage{threeparttable}
\usepackage{textcomp}
\usepackage{setspace}
\usepackage{caption}
\usepackage{comment}

\crefname{figure}{Figure}{Figures}
\crefname{table}{Table}{Tables}
\crefname{equation}{Eq.}{Eqs.}
\crefname{section}{Section}{Sections}

\SectionNumbersOn
\newcommand*{\degree}{\ensuremath{^\circ}\xspace}

\makeatletter
\let\l@addto@macro\relax
\makeatother
\usepackage[fontsize=11pt]{scrextend}

\let\oldmaketitle\maketitle
\let\maketitle\relax
\author{Carlos E.\@ V.\@ de Moura}
\email{vieirademoura.2@osu.edu}
\affiliation[The Ohio State University]
{Department of Chemistry and Biochemistry,
The Ohio State University,
Columbus, Ohio 43210, United States}
\author{Alexander Yu.\@ Sokolov}
\affiliation[The Ohio State University]
{Department of Chemistry and Biochemistry,
The Ohio State University,
Columbus, Ohio 43210, United States}
\email{sokolov.8@osu.edu}

\title{{\color{blue}
    Efficient Spin-Adapted Implementation of Multireference Algebraic Diagrammatic Construction Theory. I. Core-Ionized States and X-Ray Photoelectron Spectra
}}

\begin{document}

%%%%%%%%%%%%%%%%%%%%%%%%%%%%%%%%%%%%%%%%%%%%%%%%%%%%%%%%%%%%%%%%%%%%%
%% The "tocentry" environment can be used to create an entry for the
%% graphical table of contents. It is given here as some journals
%% require that it is printed as part of the abstract page. It will
%% be automatically moved as appropriate.
%%%%%%%%%%%%%%%%%%%%%%%%%%%%%%%%%%%%%%%%%%%%%%%%%%%%%%%%%%%%%%%%%%%%%
\begin{tocentry}
\includegraphics[width=1.0\textwidth]{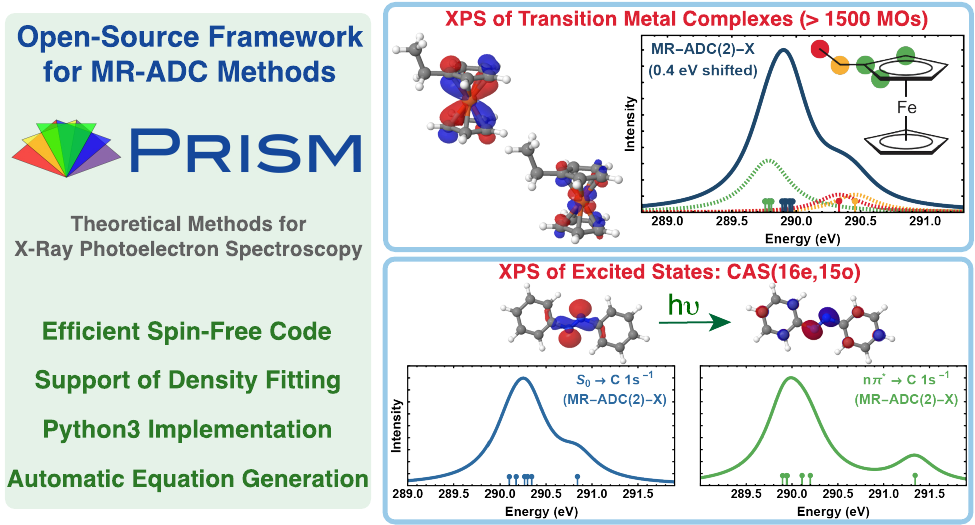}
\end{tocentry}

%%%%%%%%%%%%%%%%%%%%%%%%%%%%%%%%%%%%%%%%%%%%%%%%%%%%%%%%%%%%%%%%%%%%%
%% The abstract environment will automatically gobble the contents
%% if an abstract is not used by the target journal.
%%%%%%%%%%%%%%%%%%%%%%%%%%%%%%%%%%%%%%%%%%%%%%%%%%%%%%%%%%%%%%%%%%%%%
%ArXiv %%%%%%%%%%%%%%
\newcommand*{\abstractext}{
We present an efficient implementation of multireference algebraic diagrammatic construction theory (MR-ADC) for simulating core-ionized states and X-ray photoelectron spectra (XPS). 
Taking advantage of spin adaptation, automatic code generation, and density fitting, our implementation can perform calculations for molecules with more than 1500 molecular orbitals, incorporating static and dynamic correlation in the ground and excited electronic states. 
We demonstrate the capabilities of MR-ADC methods by simulating the XPS spectra of substituted ferrocene complexes and azobenzene isomers.
For the ground electronic states of these molecules, the XPS spectra computed using the extended second-order MR-ADC method (MR-ADC(2)-X) are in a very good agreement with available experimental results. 
We further show that MR-ADC can be used as a tool for interpreting or predicting the results of time-resolved XPS measurements by simulating the core ionization spectra of azobenzene along its photoisomerization, including the XPS signatures of excited states and the minimum energy conical intersection.
This work is the first in a series of publications reporting the efficient implementations of MR-ADC methods.
\vspace{0.25cm}
}
%ArXiv %%%%%%%%%%%%%%

%ArXiv %%%%%%%%%%%%%%
\twocolumn[
\begin{@twocolumnfalse}
\oldmaketitle
\vspace{-0.75cm}
\begin{abstract}
\abstractext
\end{abstract}
\end{@twocolumnfalse}
]
%ArXiv %%%%%%%%%%%%%%

%%%%%%%%%%%%%%%%%%%%%%%%%%%%%%%%%%%%%%%%%%%%%%%%%%%%%%%%%%%%%%%%%%%%%
%% Start the main part of the manuscript here.
%%%%%%%%%%%%%%%%%%%%%%%%%%%%%%%%%%%%%%%%%%%%%%%%%%%%%%%%%%%%%%%%%%%%%

%%%%%%%%%%%%%%%%%%%%%%%%%%%%%%%%%%%%%%%%%%%%%%%%%%%%%%%%%%%%%%%%%%%%%
%% 1. Introduction
%%%%%%%%%%%%%%%%%%%%%%%%%%%%%%%%%%%%%%%%%%%%%%%%%%%%%%%%%%%%%%%%%%%%%
\section{Introduction}
\label{sec:introduction}

%% 1.1-2. Contextualization of recent experimental advances in electronic spectroscopic methods

Understanding and harnessing light-matter interactions is a highly active area of research.
Excited electronic states populated with light are central to photochemistry, solar energy conversion, and photocatalysis where light is used to enable or accelerate chemical transformations.\cite{Zhang:2019p2216,Oeberg:2016p9631,Hoffmann:2012p1613,Balzani:2015p11320,Bonfield:2020p804,Twilton:2017p52,Sender:2017p1159,Noel:2017,Coropceanu:2019p689,Mai:2020p16832,Markushyna:2022p202200026,Srivastava:2022p100488}
Additionally, excited states play a key role in spectroscopy where their measurements provide information about the atomic and electronic structure of chemical systems.\cite{Herzberg:1966,Woerner:2013}
Recent advances in experimental techniques enable the spectroscopic studies of chemical systems in short-lived electronic states,\cite{Marek:2013p74202,GarciaRuiz:2020p396,Gaffney:2021p8010,Reinhard:2021p1086} at non-equilibrium geometries\cite{Polli:2010p440,Neville:2018p243001}, and along reaction pathways\cite{Woerner:2010p604,Ruckenbauer:2016p35522,Severino:2024p1}.

As spectroscopic techniques continue to evolve, there is a growing demand for quantum chemical methods that can accurately interpret or predict the spectral features of chemical systems with a wide range of nuclear geometries and electronic structures.
Recently, we proposed multireference algebraic diagrammatic construction theory (MR-ADC),\cite{Sokolov:2018p204113,Sokolov:2024p10.1016/bs.aiq.2024.04.004} which is a linear-response approach that allows to simulate a variety of spectroscopic properties for molecules in multiconfigurational electronic states and across potential energy surfaces.
The MR-ADC methods are similar to low-order multireference perturbation theories\cite{Andersson:1990p5483,Andersson:1992p1218,Hirao:1992p374,Finley:1998p299,Angeli:2001p259,Angeli:2001p297,Angeli:2001p10252,Angeli:2004p4043,Granovsky:2011p214113} in computational cost and can compute a variety of electronic spectra (e.g., UV/Vis and X-ray absorption,\cite{Mazin:2021p6152,Mazin:2023p4991} UV and X-ray photoelectron\cite{Chatterjee:2019p5908,Chatterjee:2020p6343,Moura:2022p4769,Moura:2022p8041}) for chemical systems with many electrons and molecular orbitals.
However, due to their inefficient spin-orbital implementation, all reported MR-ADC calculations have so far been limited to small molecules.

In this work, we demonstrate that combining spin adaptation, density fitting, and automatic code generation allows to implement MR-ADC efficiently, enabling applications to large molecules and one-electron basis sets.
Focusing on the MR-ADC methods for simulating core-ionized states and X-ray or extreme ultraviolet photoelectron spectra (XPS), this paper is the first of a series reporting fast MR-ADC implementations for a variety of spectroscopic processes.
XPS is a widely used technique for the experimental characterization of molecules and materials, measuring the element-specific core-electron binding energies that are highly sensitive to oxidation states and local chemical environment.\cite{Fadley:2010p2,Greczynski:2022p11101,Greczynski:2023p40}
The utility of XPS is further extended by the time-resolved XPS (TR-XPS) that enables studies of molecules in electronically excited states and along the course of chemical reactions.\cite{Stolow:2004p1719,Neppl:2015p64,Brausse:2018p43429,Roth:2021p1196,Mayer:2022p198,Shavorskiy,Costantini:2022p147141,Ozawa:2019p4388,Roth:2019p20303,Arion:2015p121602,Schuurman:2022p20012,Gabalski:2023p7126,Keefer:2023p73} 
However, the XPS and TR-XPS spectra usually exhibit broad overlapping features that can be difficult to interpret without insights from accurate theoretical calculations.

Here, we demonstrate the capabilities of our efficient MR-ADC implementation by simulating and interpreting the XPS spectra of substituted ferrocene complexes in their ground electronic states and the azobenzene molecule along its excited-state isomerization.
Our calculations employ large core-valence polarized basis sets correlating all electrons in up to 1532 molecular orbitals.
For the substituted ferrocene complexes, we show that the carbon K-edge XPS spectra simulated using the extended second-order MR-ADC method (MR-ADC(2)-X) are in a very good agreement with the experimental data, providing accurate interpretation of overlapping spectral features.
In our study of azobenzene, we predict the carbon and nitrogen K-edge XPS signatures of ground and excited electronic states at the {\it cis}, {\it trans}, and conical intersection geometries, which may be helpful in interpreting the results of TR-XPS experiments in the future.

Our paper is organized as follows.
First, we briefly cover the theoretical foundations of MR-ADC for simulating core-ionized states and XPS spectra and discuss its spin adaptation (\cref{sec:theory}).
We then describe the details of our efficient implementation, including the automatic generation of spin-adapted code and density fitting (\cref{sec:implementation}).
We provide the computational details in \cref{sec:comp_details}  and present the applications of MR-ADC methods to substituted ferrocene complexes and azobenzene photoisomerization in \cref{sec:results}.
Our conclusions are presented in \cref{sec:conclusions}.
Additional computational details and working equations are provided in the Supplementary Information.

%%%%%%%%%%%%%%%%%%%%%%%%%%%%%%%%%%%%%%%%%%%%%%%%%%%%%%%%%%%%%%%%%%%%%
%% 2. Theory
%%%%%%%%%%%%%%%%%%%%%%%%%%%%%%%%%%%%%%%%%%%%%%%%%%%%%%%%%%%%%%%%%%%%%
\section{Theory}
\label{sec:theory}

\subsection{Multireference algebraic diagrammatic construction theory}
\label{sec:theory:mr_adc}

Multireference algebraic diagrammatic construction (MR-ADC) simulates electronic excitations and spectra by approximating a retarded propagator (also known as a linear-response function) using multireference perturbation theory.\cite{Banerjee:2023p3037,Sokolov:2024p10.1016/bs.aiq.2024.04.004}
A retarded propagator describes the response of a chemical system in an electronic state $\ket{\Psi}$ with energy $E$ to a periodic perturbation with frequency $\omega$ and has a general form:
\begin{align}
	\label{eq:retarded_propagator}
		G_{\mu\nu}(\omega) &= G^{+}_{\mu\nu}(\omega) \pm G^{-}_{\mu\nu}(\omega) \notag \\
		&= \langle \Psi | q_\mu (\omega - H + E)^{-1} q^{\dag}_{\nu} | \Psi \rangle  \notag \\
        &\pm \langle \Psi | q^{\dag}_{\nu} (\omega + H - E)^{-1} q_{\mu} | \Psi \rangle \ ,
\end{align}
where $G^{+}_{\mu\nu}$ and $G^{-}_{\mu\nu}$  are called the forward and backward components and $H$ is the electronic (Born--Oppenheimer) Hamiltonian.
The operators $q_{\mu}$ and $q^{\dag}_{\nu}$ define the nature of periodic perturbation (e.g., electric or magnetic field), the physical observables of interest (e.g., density of states, polarization), and the sign of the second term.

Depending on the form of operators $q_{\mu}$ and $q^{\dag}_{\nu}$, MR-ADC can simulate a variety of spectroscopic processes, including electronic excitations in UV/Vis absorption spectroscopy,\cite{Sokolov:2018p204113,Mazin:2021p6152} ionization and electron attachment in photoelectron experiments,\cite{Chatterjee:2019p5908,Chatterjee:2020p6343}  and core excitations in X-ray absorption or photoelectron measurements.\cite{Moura:2022p4769,Moura:2022p8041,Mazin:2023p4991}
Similar to its single-reference counterpart,\cite{Schirmer:1982p2395,Schirmer:1991p4647,Mertins:1996p2140,Schirmer:1998p4734,Schirmer:2004p11449,Dreuw:2015p82,Banerjee:2023p3037} MR-ADC expresses \cref{eq:retarded_propagator} in a mathematical form where $G^{+}_{\mu\nu}$ and $G^{-}_{\mu\nu}$ are independently represented in terms of non-diagonal tensors
\begin{equation}
	\mathbf{G_{\pm}}(\omega) = \mathbf{T_{\pm}} (\omega \mathbf{S_{\pm}} - \mathbf{M_{\pm}})^{-1} \mathbf{T_{\pm}^{\dag}} \ ,
\end{equation}
which are called the effective Hamiltonian ($\mathbf{M_{\pm}}$), effective transition moments ($\mathbf{T_{\pm}}$), and overlap ($\mathbf{S_{\pm}}$) matrices.
Here, $\mathbf{M_{\pm}}$ and $\mathbf{T_{\pm}}$ are expressed in a basis of nonorthogonal excitations with overlap $\mathbf{S_{\pm}}$ and contain information about the energies and probabilities of electronic transitions for a specific spectroscopic process, respectively.

Expanding $\mathbf{M_{\pm}}$, $\mathbf{T_{\pm}}$, and $\mathbf{S_{\pm}}$ using multireference perturbation theory
\begin{align}
	\label{eq:M_approx}
	\mathbf{M_\pm} &\approx \mathbf{M^{(0)}_\pm} + \mathbf{M^{(1)}_\pm} + \ldots + \mathbf{M^{(n)}_\pm} \ , \\
	\label{eq:T_approx}
	\mathbf{T_\pm} &\approx \mathbf{T^{(0)}_\pm} + \mathbf{T^{(1)}_\pm} + \ldots + \mathbf{T^{(n)}_\pm} \ , \\
	\label{eq:S_approx}
	\mathbf{S_\pm} &\approx \mathbf{S^{(0)}_\pm} + \mathbf{S^{(1)}_\pm} + \ldots + \mathbf{S^{(n)}_\pm} \ ,
\end{align}
and truncating these expansions at order ($n$) defines the hierarchy of MR-ADC($n$) approximations.
To ensure that the MR-ADC($n$) methods are free from intruder-state problems,\cite{Andersson:1994p391,Evangelisti:1987p4930,Evangelista:2014p54109} the series in \cref{eq:M_approx,eq:T_approx,eq:S_approx} are generated with respect to the reference complete active space self-consistent field (CASSCF) wavefunction ($\ket{\Psi_0}$) and the perturbation operator $V = H - H^{(0)}$, where $H^{(0)}$ is the Dyall zeroth-order Hamiltonian.\cite{Dyall:1995p4909,Sokolov:2024p10.1016/bs.aiq.2024.04.004}

The MR-ADC($n$) excitation energies relative to the reference state are computed as the eigenvalues of $\mathbf{M_{\pm}}$ ($\boldsymbol{\Omega_\pm}$) by solving the Hermitian generalized eigenvalue problem
\begin{equation}
	\label{eq:mradc-eigenvalue-problem}
	\mathbf{M_\pm} \mathbf{Y_\pm} = \mathbf{S_\pm} \mathbf{Y_\pm} \boldsymbol{\Omega_\pm} \ .
\end{equation}
The eigenvectors $\mathbf{Y_\pm}$ are used to compute the spectroscopic amplitudes
\begin{equation}
	\label{eq:mradc-spectroscopic-amplitudes}
	\mathbf{X_\pm} = \mathbf{T_\pm} \mathbf{S_\pm^{-1/2}} \mathbf{Y_\pm}
\end{equation}
that provide access to transition intensities, densities of states, and spectra.
When expressed in the eigenstate basis of $\mathbf{M_\pm}$, the MR-ADC($n$) propagator can be written as
\begin{equation}
	\label{eq:ADC_diag}
	\mathbf{G_\pm}(\omega) = \mathbf{X_\pm} (\omega\mathbf{1} - \mathbf{\Omega_\pm})^{-1}  \mathbf{X_\pm^\dag} \ ,
\end{equation}
which is known as its spectral representation.

\subsection{MR-ADC for core-ionized states and X-ray photoelectron spectra}
\label{sec:theory:mr_adc_xps}

\begin{figure*}[t!]
	\centering
	\includegraphics[width=1.0\textwidth]{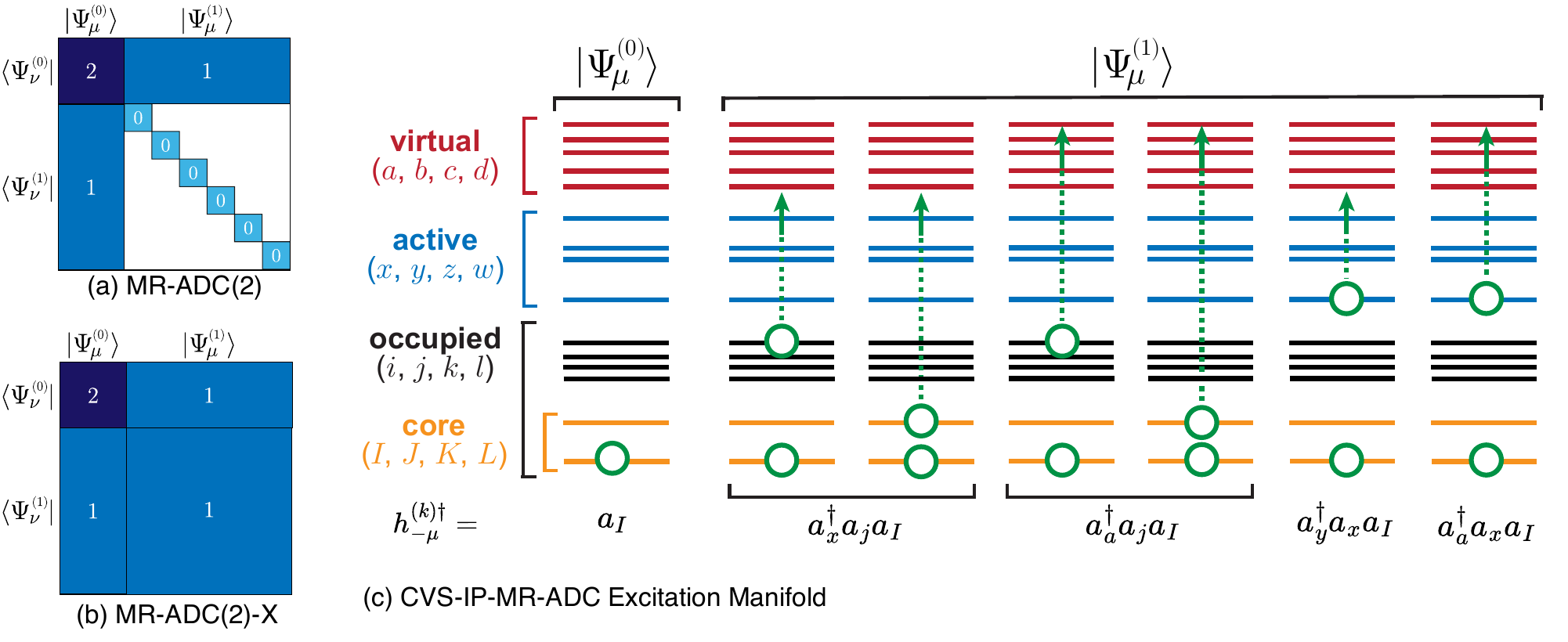}
	\caption{
		Schematic representation of the effective Hamiltonian matrix $\mathbf{M}_-$ for the (a) CVS-IP-MR-ADC(2) and (b) CVS-IP-MR-ADC(2)-X approximations in the basis of electronic excitations $\ket{\Psi_\mu^{(k)}} =  h_{-\mu}^{(k)\dag}\ket{\Psi_0}$ (c).
		Nonzero matrix blocks are highlighted in color.
		Numbers represent the perturbation order in which the effective Hamiltonian is evaluated in each matrix block.
		}
	\label{fig:cvs_mradc_matrix_compare}
\end{figure*}

In this work, we will focus on the MR-ADC methods for simulating core-ionized states and X-ray photoelectron spectra, which are derived from the backward component of one-particle Green's function\cite{Fetter2003,Dickhoff2005}
\begin{equation}
	\label{eq:1-gf_ip}
		G_{-pq}(\omega)
		= \langle \Psi | a_q^{\dag} (\omega + H - E)^{-1} a_p | \Psi \rangle \ .
\end{equation}
Approximating $\mathbf{G_{-}}(\omega)$ in \cref{eq:1-gf_ip} following the approach described in \cref{sec:theory:mr_adc} gives rise to the MR-ADC($n$)  approximations that describe the $(N-1)$-electron ionized states starting from the $N$-electron CASSCF wavefunction (IP-MR-ADC).\cite{Chatterjee:2019p5908,Chatterjee:2020p6343}
In contrast to conventional multireference perturbation theories for excited states,\cite{Andersson:1990p5483,Andersson:1992p1218,Hirao:1992p374,Finley:1998p299,Angeli:2001p259,Angeli:2001p297,Angeli:2001p10252,Angeli:2004p4043,Granovsky:2011p214113} the IP-MR-ADC methods do not require separate calculations for the reference and ionized states, and can simulate excitations in all (active and non-active) molecular orbitals.
These features make IP-MR-ADC particularly attractive for calculating the excited states and spectra measured in X-ray photoelectron spectroscopy (XPS) experiments where an electron is ejected from a core or inner-shell valence orbital following the excitation with X-ray or extreme ultraviolet light.
However, computing these high-energy states directly by solving the generalized eigenvalue problem in \cref{eq:mradc-eigenvalue-problem} is very difficult as they are deeply embedded in the eigenstate spectrum of IP-MR-ADC effective Hamiltonian matrix $\mathbf{M_-}$.

To overcome this challenge, IP-MR-ADC is combined with a core-valence separation (CVS)\cite{Cederbaum:1980p206,Barth:1981p1038,Angonoa:1987p6789,Schirmer:2001p10621,Thiel:2003p2088} approximation that neglects the coupling between valence- and core-ionized states due to their large difference in energy and spatial localization.\cite{Moura:2022p4769,Moura:2022p8041}
In the CVS-IP-MR-ADC methods, the molecular orbitals of the system are split into four subsets: ``core'', ``valence'', ``active'', and ``virtual'' (\cref{fig:cvs_mradc_matrix_compare}).
The union of core and valence subspaces corresponds to the doubly occupied non-active orbitals in the reference CASSCF wavefunction $\ket{\Psi_0}$.
The core subspace includes all lowest-energy molecular orbitals starting with the one that is expected to be ionized first in the XPS spectrum.
The remaining non-active occupied orbitals are incorporated in the valence subspace.
The CVS-IP-MR-ADC matrices ($\mathbf{M_-}$, $\mathbf{T_-}$, and $\mathbf{S_-}$) are expressed in the basis of ionized electronic configurations $\ket{\Psi_\mu^{(k)}} =  h_{-\mu}^{(k)\dag}\ket{\Psi_0}$ where the $k$th-order excitation operators $h_{-\mu}^{(k)\dag}$ are required to ionize or excite an electron from at least one core orbital (\cref{fig:cvs_mradc_matrix_compare}).

The $n$th-order contributions to $\mathbf{M_-}$, $\mathbf{T_-}$, and $\mathbf{S_-}$ have the form:
\begin{align}
	\label{eq:M-_matrix}
	M_{-\mu\nu}^{(n)} &= \sum_{klm}^{k+l+m=n} \braket{\Psi_0|[h_{-\mu}^{(k)\dagger},[\tilde{H}^{(l)},h_{-\nu}^{(m)}]]_{+}|\Psi_0} \ , \\
	\label{eq:T-_matrix}
	T_{-p\nu}^{(n)} &= \sum_{kl}^{k+l=n} \braket{\Psi_0|[\tilde{a}_p^{(k)},h_{-\nu}^{(l)}]_{+}|\Psi_0} \ , \\
	\label{eq:S-_matrix}
	S_{-\mu\nu}^{(n)} &= \sum_{kl}^{k+l=n} \braket{\Psi_0|[h_{-\mu}^{(k)\dagger},h_{-\nu}^{(l)}]_{+}|\Psi_0} \ ,
\end{align}
where $\tilde{H}^{(k)}$ and $\tilde{a}_p^{(k)}$ are the $k$th-order components of effective Hamiltonian and effective observable operators, $[A,B] = AB - BA$ and  $[A,B]_+ = AB + BA$ denote commutator and anticommutator, respectively.\cite{Sokolov:2018p204113, Chatterjee:2019p5908, Chatterjee:2020p6343, Sokolov:2024p10.1016/bs.aiq.2024.04.004}
\cref{eq:M-_matrix,eq:T-_matrix,eq:S-_matrix} define the perturbative structure of CVS-IP-MR-ADC matrices, as exemplified in \cref{fig:cvs_mradc_matrix_compare} for the $\mathbf{M_-}$ of strict second-order (CVS-IP-MR-ADC(2)) and extended second-order (CVS-IP-MR-ADC(2)-X) approximations that will be employed in this work.\cite{Moura:2022p4769,Moura:2022p8041}
Both methods incorporate the single  ($\ket{\Psi_\mu^{(0)}}$) and double ($\ket{\Psi_\mu^{(1)}}$) excitations out of the reference wavefunction $\ket{\Psi_0}$ (described by the $h_{-\mu}^{(0)\dag}$ and $h_{-\mu}^{(1)\dag}$ operators, respectively), and expand the effective Hamiltonian $\tilde{H}^{(k)}$ to the second order in the $\bra{\Psi_\nu^{(0)}} - \ket{\Psi_\mu^{(0)}}$ block and to the first order for the $\bra{\Psi_\nu^{(1)}} - \ket{\Psi_\mu^{(0)}}$ and $\bra{\Psi_\nu^{(0)}} - \ket{\Psi_\mu^{(1)}}$ sectors.
The CVS-IP-MR-ADC(2)-X method provides a higher-level description of correlation effects in the $\bra{\Psi_\nu^{(1)}} - \ket{\Psi_\mu^{(1)}}$ block by including the contributions from $\tilde{H}^{(1)}$, which significantly improve the description of excited-state orbital relaxation effects.

We note that the CVS approximation used in this work is different from the one originally proposed in the context of single-reference ADC methods\cite{Angonoa:1987p6789,Schirmer:2001p10621,Thiel:2003p2088} where double excitations from two core orbitals were excluded. 
Including these doubly excited configurations has been shown to significantly improve the accuracy of CVS approximation\cite{Peng:2019p1840,Brumboiu:2022p214109,Herbst:2020p54114} and has been widely used in other electronic structure methods based on CVS\cite{Wenzel:2014p1900,Coriani:2015p181103,Liu:2019p1642,Zheng:2019p4945,Vidal:2019p3117,Simons:2022p3759,Zheng:2022p13587}.

In our earlier work,\cite{Chatterjee:2020p6343,Moura:2022p4769} the CVS-IP-MR-ADC(2) and CVS-IP-MR-ADC(2)-X methods were implemented by deriving  \cref{eq:M-_matrix,eq:T-_matrix,eq:S-_matrix} in the basis of spin-orbitals $\psi_p (1) = \phi_p(1) \sigma_p(1)$ where $\phi_p(1)$ and $\sigma_p(1)$ are the spatial and spin components of $\psi_p (1)$.
The resulting working equations can be fully expressed in terms of the one- and antisymmetrized two-electron integrals ($h_p^q$, $\braket{pq||rs}$), the energies of non-active orbitals ($\epsilon_p$), the correlation (cluster) amplitudes of effective Hamiltonian and observable operators ($t_{p}^{q(k)}$, $t_{pq}^{rs(k)}$), and the reduced density matrices of reference wavefunction $\ket{\Psi_0}$ ($\gamma^p_q = \braket{\Psi_0|a_p^\dag a_q|\Psi_0}$, $\gamma^{pq}_{rs} = \braket{\Psi_0|a_p^\dag a_q^\dag a_s a_r|\Psi_0}$, etc.).
However, due to the large computational costs associated with computing, storing, and contracting spin-orbital tensors, the CVS-IP-MR-ADC(2) and CVS-IP-MR-ADC(2)-X calculations were limited to chemical systems with $\lesssim$ 300 spatial orbitals.
In \cref{sec:theory:spin-free}, we describe how these bottlenecks can be avoided by formulating CVS-IP-MR-ADC(2) and CVS-IP-MR-ADC(2)-X in the spin-free basis.

\subsection{Spin-free formulation of CVS-IP-MR-ADC}
\label{sec:theory:spin-free}

When using a non-relativistic Hamiltonian, it is possible and computationally advantageous to remove the dependence on spin in equations through so-called spin adaptation.
One of the most widely employed approaches to spin-adapt multireference theories\cite{Shamasundar:2009p174109,Kutzelnigg:2010p433} is to express all operators in terms of spin-free unitary group generators (e.g., $E^{pq}_{rs} = \sum_{\rho \sigma}^{\alpha,\beta} a_{p\sigma}^\dag a_{q\rho}^\dag a_{s\rho} a_{r\sigma}$) that are invariant under the SU(2) transformations of spin-orbitals $\psi_{p\alpha} \leftrightarrow \psi_{p\beta}$ where $\alpha$ and $\beta$ indicate the up and down spin, respectively.
Once the operators are defined, their matrix elements are derived using the spin-free formulation of Wick's theorem, yielding fully spin-adapted equations.\cite{Kutzelnigg:1997p432}
While this approach can be straightforwardly used in state-specific multireference methods, it is less convenient for multistate effective Hamiltonian theories such as MR-ADC where the excited-state spin symmetry or particle number can be different from that of the reference electronic state.

In this work, we employ the spin adaptation approach developed by Kutzelnigg, Shamasundar, and Mukherjee\cite{Kutzelnigg:1999p2800,Kutzelnigg:2002p4787,Shamasundar:2009p174109,Kutzelnigg:2010p433} (KSM) that allows to eliminate spin variables from spin-orbital equations {\it a posteriori}, by utilizing the relationships between spin-free and spin-dependent tensors for the $M_S = 0$ reference state.
Importantly, the resulting equations can be used to perform spin-adapted calculations starting with a closed- ($S = 0$) or open-shell ($S > 0$) reference state, as long as the reference wavefunction is an equally weighted ensemble of the entire spin multiplet ($\{ \ket{\Psi_{0}^{S, M_S}}\}$, $M_S$ $=$ $-S, \ldots, S$).
Such reference state can be computed using the state-averaged CASSCF method, which is available in many quantum chemistry software packages.
In addition to its straightforward implementation, the KSM approach enforces the $M_S$ degeneracy of computed open-shell excited states, which can be violated in calculations with a pure-state open-shell reference.
The KSM spin adaptation was successfully used to develop efficient implementations of several effective Hamiltonian theories, such as state-specific partially internally contracted multireference coupled cluster theory,\cite{Datta:2011p214116,Datta:2013p2639} multireference equation-of-motion coupled cluster theory,\cite{Datta:2012p204107,Huntington:2016p114} anti-Hermitian contracted Schr\"odinger equation,\cite{Mazziotti:1998p4219,Mazziotti:2006p32501,Mazziotti:2006p143002,Mazziotti:2007p22505} and multireference driven similarity renormalization group.\cite{Li:2021p114111}

Starting with the $M_S = 0$ reference wavefunction ($\ket{\Psi_{0}}$), we spin-adapt $\mathbf{M_-}$, $\mathbf{T_-}$, and $\mathbf{S_-}$ (\cref{eq:M-_matrix,eq:T-_matrix,eq:S-_matrix}) using the following relationships for the non-active orbital energies ($\epsilon_p$), the one- and antisymmetrized two-electron integrals ($h_p^q$, $\braket{pq||rs}$), and the amplitudes of effective operators $\tilde{H}^{(k)}$ and $\tilde{a}_p^{(k)}$ ($t_{p}^{q(k)}$, $t_{pq}^{rs(k)}$):
\begin{equation}
	\epsilon_{p_{\alpha}} = \epsilon_{p_{\beta}} = \epsilon_{p} \ ,
\end{equation}
\begin{equation}
	h_{p_{\alpha}}^{q_{\alpha}} = h_{p_{\beta}}^{q_{\beta}} = h_{p}^{q} \ ,
\end{equation}
\begin{equation}
	\braket{p_{\alpha} q_{\beta}||r_{\alpha} s_{\beta}} = \braket{p_{\beta} q_{\alpha}||r_{\beta} s_{\alpha}} = v_{pq}^{rs} \ ,
\end{equation}
\begin{equation}
	\braket{p_{\alpha} q_{\alpha}||r_{\alpha} s_{\alpha}} = \braket{p_{\beta} q_{\beta}||r_{\beta} s_{\beta}} = v_{pq}^{rs} - v_{pq}^{sr} \ ,
\end{equation}
\begin{equation}
	t_{p_{\alpha}}^{q_{\alpha}(k)} = t_{p_{\beta}}^{q_{\beta}(k)} = t_{p}^{q(k)} \ ,
\end{equation}
\begin{equation}
	t_{p_{\alpha} q_{\beta}}^{r_{\alpha} s_{\beta}(k)} = t_{p_{\beta} q_{\alpha}}^{r_{\beta} s_{\alpha}(k)} = t_{pq}^{rs(k)} \ ,
\end{equation}
\begin{equation}
	t_{p_{\alpha} q_{\alpha}}^{r_{\alpha} s_{\alpha}(k)} = t_{p_{\beta} q_{\beta}}^{r_{\beta} s_{\beta}(k)} = t_{pq}^{rs(k)} - t_{pq}^{sr(k)} \ ,
\end{equation}
where the r.h.s.\@ of each equation is written in terms of spin-free tensors $\epsilon_{p}$, $h_{p}^{q}$, $v_{pq}^{rs}$, $t_{p}^{q(k)}$, and $t_{pq}^{rs(k)}$.

Additionally, $\mathbf{M_-}$, $\mathbf{T_-}$, and $\mathbf{S_-}$ depend on up to four-particle active-space reduced density matrices of $\ket{\Psi_{0}}$ ($n$-RDM, $1\le n \le 4$).
The relationships between spin-orbital ($\gamma$) and spin-free ($\Gamma$) RDMs up to $n = 3$ have been previously reported.\cite{Kutzelnigg:2010p433}
For the 1- and 2-RDMs, these equations can be written as:
\begin{equation}
	\gamma^{p_\alpha}_{q_\alpha} = \gamma^{p_\beta}_{q_\beta} = \frac{1}{2} \Gamma^{p}_{q} \ ,
\end{equation}
\begin{equation}
	\gamma^{p_\alpha q_\alpha}_{r_\alpha s_\alpha} = \gamma^{p_\beta q_\beta}_{r_\beta s_\beta} = \frac{1}{6} \Gamma^{pq}_{rs} - \frac{1}{6} \Gamma^{pq}_{sr} \ ,
\end{equation}
\begin{equation}
	\gamma^{p_\alpha q_\beta}_{r_\alpha s_\beta} = \gamma^{p_\beta q_\alpha}_{r_\beta s_\alpha} = \frac{1}{3} \Gamma^{pq}_{rs} + \frac{1}{6} \Gamma^{pq}_{sr} \ ,
\end{equation}
\begin{equation}
	\gamma^{p_\alpha q_\beta}_{r_\beta s_\alpha} = \gamma^{p_\beta q_\alpha}_{r_\alpha s_\beta} = - \left( \frac{1}{6} \Gamma^{pq}_{rs} + \frac{1}{3} \Gamma^{pq}_{sr} \right) \ .
\end{equation}
For $n = 3$, these relationships are rather complicated and can be found in the Supplementary Information.\cite{Kutzelnigg:2010p433}
To the best of our knowledge, the equations for spin-adapting 4-RDM have not been published.
We derived them using the approach outlined in Ref.\@ \citenum{Kutzelnigg:2010p433} and included in the Supplementary Information.
We note that the IP-MR-ADC(2) and IP-MR-ADC(2)-X methods can be implemented without 4-RDM by factorizing its contributions into intermediates,\cite{Chatterjee:2020p6343} although we have not taken advantage of this yet in our efficient implementation.

Finally, the CVS-IP-MR-ADC(2) and CVS-IP-MR-ADC(2)-X equations can be further simplified by taking advantage of spin symmetry in the excitation manifold $h_{-\mu}^{(k)\dag}\ket{\Psi_0}$ (\cref{fig:cvs_mradc_matrix_compare}) for the $M_S = 0$ reference state. 
For example, when calculating the matrix-vector products $\boldsymbol{\sigma_-} = \mathbf{M_{-}} \mathbf{Z_{-}}$ for an arbitrary vector $\mathbf{Z_{-}}$, the $\sigma_{-\mu}$ matrix elements need to be evaluated only for the following excitations:
\begin{equation}
	h_{-\mu}^{(0)\dag} \in \{ a_{I_{\alpha}} \} \ ,
\end{equation}
\begin{align}
	h_{-\mu}^{(1)\dag} \in
	         \{ &a_{I_{\alpha} J_{\alpha}}^{x_{\alpha}},
				 a_{I_{\alpha} J_{\beta}}^{x_{\beta}};
				 a_{I_{\alpha} j_{\alpha}}^{x_{\alpha}},
				 a_{I_{\alpha} j_{\beta}}^{x_{\beta}},
				 a_{I_{\beta}  j_{\alpha}}^{x_{\beta}}; \notag \\
				 &a_{I_{\alpha} J_{\alpha}}^{a_{\alpha}},
				 a_{I_{\alpha} J_{\beta}}^{a_{\beta}};
				 a_{I_{\alpha} j_{\alpha}}^{a_{\alpha}},
				 a_{I_{\alpha} j_{\beta}}^{a_{\beta}},
				 a_{I_{\beta}  j_{\alpha}}^{a_{\beta}}; \notag \\
				 &a_{I_{\alpha} x_{\alpha}}^{y_{\alpha}},
				 a_{I_{\alpha} x_{\beta}}^{y_{\beta}},
				 a_{I_{\beta}  x_{\alpha}}^{y_{\beta}};
				 a_{I_{\alpha} x_{\alpha}}^{a_{\alpha}},
				 a_{I_{\alpha} x_{\beta}}^{a_{\beta}},
				 a_{I_{\beta}  x_{\alpha}}^{a_{\beta}} \} \ ,
\end{align}
where we used the index notation from \cref{fig:cvs_mradc_matrix_compare}c and denoted $a^{r}_{pq} \equiv a_r^\dag a_q a_p$ for brevity.
The remaining matrix elements can be obtained, if necessary, by utilizing the spin and permutational symmetry of $h_{-\mu}^{(k)\dag}\ket{\Psi_0}$.

%%%%%%%%%%%%%%%%%%%%%%%%%%%%%%%%%%%%%%%%%%%%%%%%%%%%%%%%%%%%%%%%%%%%%
%% 3. Implementation
%%%%%%%%%%%%%%%%%%%%%%%%%%%%%%%%%%%%%%%%%%%%%%%%%%%%%%%%%%%%%%%%%%%%%
\section{Implementation}
\label{sec:implementation}

\subsection{Overview}
\label{sec:implementation:overview}

The CVS-IP-MR-ADC(2) and CVS-IP-MR-ADC(2)-X methods were implemented using Python in the open-source and freely available program Prism.\cite{Prism}
To obtain the CASSCF orbitals, one- and two-electron integrals, and the reference reduced density matrices, Prism was interfaced with the PySCF package.\cite{Sun:2020p24109}
The CVS-IP-MR-ADC code follows an algorithm summarized below:
\begin{enumerate}
  \item Perform the reference CASSCF calculation for a specified molecular geometry, basis set, and active space using PySCF.
  Compute the occupied and virtual orbital energies as the eigenvalues of generalized Fock matrix and transform the one- and two-electron integrals to the molecular orbital basis.
  \item Compute the $t_{p}^{q(1)}$, $t_{p}^{q(2)}$, and $t_{pq}^{rs(1)}$ amplitudes of effective Hamiltonian by solving the amplitude equations as discussed in Ref.\@ \citenum{Chatterjee:2020p6343}.
  The $t_{p}^{q(1)}$ and $t_{pq}^{rs(1)}$ amplitudes parameterize the first-order wavefunction in fully internally contracted second-order N-electron valence perturbation theory (fic-NEVPT2)\cite{Angeli:2001p10252,Angeli:2001p297} and are used to compute the fic-NEVPT2 correlation energy for the reference state.
  \item Evaluate the  $\mathbf{M_-}$ matrix elements in the $\bra{\Psi_\nu^{(0)}} - \ket{\Psi_\mu^{(0)}}$ ($\mathbf{M_-^{00}}$) and $\bra{\Psi_\nu^{(1)}} - \ket{\Psi_\mu^{(0)}}$ ($\mathbf{M_-^{10}}/\mathbf{M_-^{01}}$) sectors (\cref{fig:cvs_mradc_matrix_compare}).
  Since the number of single excitations ($\ket{\Psi_\mu^{(0)}}$) is very small (equal to the user-defined number of core orbitals), the $\mathbf{M_-^{00}}$ and $\mathbf{M_-^{10}}/\mathbf{M_-^{01}}$ blocks are stored in memory as reusable intermediates for the rest of the calculation.
  \item Solve the eigenvalue problem in \cref{eq:mradc-eigenvalue-problem} by iteratively optimizing the eigenvectors $\mathbf{Y_-}$ using the multiroot Davidson algorithm\cite{Davidson:1975p87} for the requested number of lowest-energy excited states.
  \item From the converged eigenvectors $\mathbf{Y_-}$ compute the spectroscopic amplitudes $\mathbf{X_-}$ (\cref{eq:mradc-spectroscopic-amplitudes}) and transition intensities.
\end{enumerate}

Our efficient implementation of CVS-IP-MR-ADC(2) and CVS-IP-MR-ADC(2)-X has several features:
(i) spin adaption of all tensor contractions (\cref{sec:theory:spin-free}) assisted by automatic equation and code generation as discussed in \cref{sec:implementation:code_generation};
(ii) efficient in-core and out-of-core handling of two-electron integrals utilizing density fitting (\cref{sec:implementation:df}) and h5py library;\cite{Collette2013}
(iii) optimized implementation of tensor contractions using basic linear algebra subroutines (BLAS), opt$\_$einsum\cite{Smith:2018p753} and numpy\cite{Harris:2020p357} modules;
and (iv) open multiprocessing (OpenMP) parallelization of computationally intensive tasks.
In the following, we provide additional details on automatic equation and code generation (\cref{sec:implementation:code_generation}) and the use of density fitting (\cref{sec:implementation:df}).

\subsection{Automatic Derivation of Spin-Free Equations and Code Generation}
\label{sec:implementation:code_generation}

The CVS-IP-MR-ADC(2)  and CVS-IP-MR-ADC(2)-X equations are algebraically complicated, which makes their manual implementation tedious and time-consuming.
For example, in the spin-orbital basis, the equations for $\mathbf{M_-^{00}}$ contain $\sim$ 200 terms.
Due to the lower permutational symmetry of spin-free tensors, the number of spin-adapted $\mathbf{M_-^{00}}$ terms reaches $\sim$ 1000.
In general, spin adaptation increases the number of tensor contractions by a factor of 3 to 5, relative to the spin-orbital implementation.
As a result, implementing CVS-IP-MR-ADC(2)  and CVS-IP-MR-ADC(2)-X efficiently requires assistance from computer software that can derive spin-adapted equations and translate them into optimized code.

Computer-aided derivation and code generation enabled many significant advances in quantum chemistry.
Widespread adoption of second quantization formalism\cite{Dirac:1927p243,Feynman:1998} created a need for tools that can handle tedious operations, such as permuting strings of creation and annihilation operators, normal ordering, evaluating commutators, and matrix elements.
Automatic derivation of second-quantized equations was pioneered by Janssen and Schaefer in the framework of coupled cluster theory (CC).\cite{Janssen:1991p1}
Since then, implementations of many single-\cite{Li:1994p8812,Kallay:2001p2945,Hirata:2003p9887,Kallay:2004p9257,Hirata:2004p51,Kallay:2005p214105,Auer:2006p211,Kamiya:2006p74111,Kamiya:2007p134112,Shiozaki:2008p3358,Koehn:2008p201103,Kvaal:2012p194109,Kvaal:2013p1100,Nakajima:2015p349,Pedersen:2019p144106,Kjellgren:2019p124113,Sumita:2021p26746,Leitner:2022p184101} and multireference methods\cite{Das:2010p74103,Hanauer:2011p204111,Datta:2012p204107,Datta:2013p2639,Saitow:2013p44118,Kim:2021p113187,Guo:2021p214113,Black:2023p134801} have been developed with the help of computer tools for equation and code generation, such as tensor contraction engine (TCE),\cite{Hirata:2003p9887} SecondQuantizationAlgebra (SQA),\cite{Neuscamman:2009p124102} SMITH,\cite{Shiozaki:2008p3358} SMITH3,\cite{MacLeod:2015p51103} ORCA-AGE,\cite{Krupicka:2017p1853} sympy,\cite{Meurer:2017p103} $p^\dag q$,\cite{Rubin:2021p1954709} WICK\&D,\cite{Evangelista:2022p64111} and QCMATH.\cite{QuinteroMonsebaiz:2023p85035,Monino:2023p}

In our spin-orbital implementation of MR-ADC methods, we utilized the SQA program\cite{Neuscamman:2009p124102} to derive expressions for the elements of $\mathbf{M_\pm}$, $\mathbf{T_\pm}$, $\mathbf{S_\pm}$, and their matrix-vector products.\cite{Chatterjee:2019p5908,Mazin:2021p6152,Chatterjee:2020p6343}
In this work, we developed an extension of SQA called SQA+\cite{sqaplus} that allows to perform spin adaptation using the KSM approach discussed in \cref{sec:theory:spin-free}.
In SQA+, every operator is initially represented in a spin-orbital form with separate labels for molecular orbitals and electron spin.
The spatial orbitals are classified according to one of the four subspaces shown in \cref{fig:cvs_mradc_matrix_compare}.
Once an operator expression is defined by the user, its matrix elements are evaluated in the spin-orbital form by normal-ordering $a_{p\sigma}^\dag$ and $a_{p\sigma}$ ($\sigma \in \{\alpha, \beta\}$) with respect to $\ket{\Psi_0}$\cite{Chatterjee:2019p5908} and discarding vanishing terms.
The spin-orbital tensor contractions are converted to the spin-free form using the KSM relationships from \cref{sec:theory:spin-free} and Supplementary Information, and are simplified by combining identical terms.
The resulting expressions are converted into Python code, which can be directly incorporated into Prism.
Additionally, SQA+ allows to rewrite the spin-free equations in terms of automatically defined reusable intermediates,\cite{Mazin:2021p6152} which can be used to make the code more efficient. 
We note that an implementation of spin-free unitary group generators in SQA has been developed by Saitow et al.,\cite{Saitow:2013p44118} although we did not employ it in our work.

\subsection{Density Fitting}
\label{sec:implementation:df}

To enable the CVS-IP-MR-ADC calculations with large one-electron basis sets, our efficient implementation utilized the density fitting (DF) approximation\cite{Whitten:1973p4496,Dunlap:1979p3396,Vahtras:1993p514,Feyereisen:1993p359,Dunlap:2000p2113}
\begin{equation}
	v_{pq}^{rs} = (pr|qs) \approx \sum_{Q}^{N_{\mathrm{aux}}} b_{pr}^{Q} b_{qs}^{Q} \ ,
\end{equation}
where the two-electron integrals $(pr|qs)$ in Chemists' notation are expressed as a product of three-index tensors $b_{pq}^{Q}$ evaluated in the basis of spatial molecular orbitals ($\phi_p$) and auxiliary basis functions ($\chi_Q$):
\begin{align}
	b_{pq}^{Q} &= \sum_{P}^{N_{aux}}( pq | P ) (J^{-\frac{1}{2}} )_{PQ}\label{eq: b_pq} \ , \\
	( pq | P ) &=  \iint \phi_{p} (1) \phi_{q}(1) \frac{1}{r_{12}} \chi_{P}(2) dr_{1} dr_{2}\label{3_ind_int} \ ,\\
	J_{PQ} &= \iint \chi_{P} (1) \frac{1}{r_{12}}\chi_{Q}(2)dr_{1} dr_{2}\label{eq: j_PQ} \ .
\end{align}
The DF approximation lowers the cost of transforming two-electron integrals from atomic to molecular basis and allows to avoid storing $v_{pq}^{rs}$ on disk, which becomes prohibitively expensive in calculations with more than 1000 molecular orbitals.
Density fitting has been widely used to reduce the cost of excited-state electronic structure methods.\cite{Haettig:2000p5154,Haettig:2005p37,Gyoerffy:2013p104104,Epifanovsky:2013p134105,Helmich:2014p35,Kumar:2017p1611,Mester:2018p94111,Herbst:2020p1462,Liu:2020p174109,Banerjee:2021p74105}
Here, we employ it for the first time in the implementation of MR-ADC.

To assess the accuracy of DF approximation in the CVS-IP-MR-ADC calculations, we computed core ionization energies for molecules from the benchmark set of Liu {et al.}\cite{Liu:2006p44102}.
The results presented in the Supplementary Information demonstrate that using the exact $v_{pq}^{rs}$ in the reference CASSCF calculations and DF-approximated two-electron integrals in CVS-IP-MR-ADC has a very small effect on core ionization energies with mean absolute error (MAE)  of $0.0002$ eV.
Introducing the DF approximation in the CASSCF step increases the MAE to 0.0011 eV.
Overall, these results suggest that DF can be used to significantly lower the cost of CVS-IP-MR-ADC calculations  without compromising their accuracy.

%%%%%%%%%%%%%%%%%%%%%%%%%%%%%%%%%%%%%%%%%%%%%%%%%%%%%%%%%%%%%%%%%%%%%
%% 4. Computational Details
%%%%%%%%%%%%%%%%%%%%%%%%%%%%%%%%%%%%%%%%%%%%%%%%%%%%%%%%%%%%%%%%%%%%%
\section{Computational Details}
\label{sec:comp_details}

In order to demonstrate the capabilities of CVS-IP-MR-ADC methods, we performed the calculations of core-ionized states and XPS spectra for four molecules: ethyl-ferrocene (EtFC), vinyl-ferrocene (VFC), ethynyl-ferrocene (EFC), and azobenzene.
For azobenzene, three structures were investigated: the equilibrium geometries of \textit{trans}- and \textit{cis}-isomers, and the geometry of minimum energy conical intersection (MECI) between the two lowest-energy singlet states ($S_0$ and $S_1$) that is important in the \textit{trans}-\textit{cis} photoisomerization.
For brevity, we do not include CVS-IP in the abbreviation of MR-ADC methods henceforth.

The ground-state geometries of all molecules were optimized using density functional theory (DFT)\cite{Hohenberg:1964p864,Kohn:1965p1133,Parr1994,Teale:2022p28700} with the B3LYP hybrid exchange-correlation functional\cite{Vosko:1980p1200,Lee:1988p785,Becke:1993p5648,Stephens:1994p11623} and the D3(BJ) dispersion correction.\cite{Grimme:2011p1456}
To compute the $S_0$--$S_1$ MECI geometry of azobenzene, we performed state-averaged complete active space self-consistent field\cite{Werner:1980p2342,Werner:1981p5794,Knowles:1985p259} calculation with equal weights for $S_0$ and $S_1$ (SA2-CASSCF). 
All geometry optimization calculations were performed using the cc-pVQZ basis set with JK density fitting\cite{Weigend:2002p3175,Weigend:2002p4285,Bross:2013p94302} in the Molpro\cite{Werner:2020p144107,Werner:2012p242,Werner} package.

Reference wavefunctions for the MR-ADC calculations were computed using CASSCF implemented in the PySCF package.\cite{Sun:2020p24109}
For EtFC, active space consisted of 8 electrons in 8 orbitals (CAS(8e,8o)) with significant contributions from the $\pi^\ast$ orbitals of cyclopentadienyl rings and the d$_\pi$ and d$_\delta$ orbitals of Fe.
In the case of VFC and EFC, one additional bonding and antibonding orbital for each $\pi$-bond in the substituent were included,  resulting in the CAS(10e,10o) and CAS(12e,12o) active spaces, respectively.
All azobenzene calculations were performed using CAS(16e,15o), which incorporated the $\pi$ and $\pi^\ast$ orbitals of phenyl rings, the \ch{N-N} $\pi$ and $\pi^\ast$ orbitals, and one occupied orbital representing nitrogen lone pairs.
The active orbitals for all molecules are visualized in the Supporting Information.
In addition to the ground electronic state, the CASSCF and MR-ADC calculations of azobenzene were performed for the $n\pi^\ast$ excited state at the \textit{trans} ($C_{2h}$ symmetry) and \textit{cis} ($C_2$) geometries labeled as $1^1B_u$ and $1^1B$, respectively.
For the \textit{trans}-isomer, we also considered the $\pi\pi^\ast$ excited state of $1^1B_g$ symmetry.
For the MECI geometry ($C_1$ symmetry), the MR-ADC calculations were performed for the $S_0$ and $S_1$ reference wavefunctions obtained from the SA2-CASSCF calculation with equal weights for each electronic state. 

The core ionization energies and XPS spectra were computed using the efficient implementation of MR-ADC in the Prism program.\cite{Prism}
As discussed in \cref{sec:implementation}, Prism was interfaced with PySCF to obtain the CASSCF orbitals, one- and two-electron integrals, and reference reduced density matrices.
All MR-ADC calculations employed the correlation-consistent core-polarized cc-pwCVQZ basis set for carbon and nitrogen\cite{Dunning:1989p1007,Peterson:2002p10548,Balabanov:2005p64107} and the cc-pVQZ basis for hydrogen and iron.
Density fitting was used for the reference CASSCF and excited-state MR-ADC methods utilizing the cc-pwCVQZ-JKFIT\cite{Weigend:2002p3175,Bross:2013p94302} and cc-pwCVQZ-RIFIT\cite{Weigend:2002p3175,Haettig:2004p59} auxiliary basis sets, respectively.
The MR-ADC calculations were performed using the $\eta_s = 10^{-6}$  and $\eta_d = 10^{-10}$ parameters to remove linearly dependent semiinternal and double excitations, respectively.\cite{Chatterjee:2019p5908}
Scalar relativistic effects were incorporated using the spin-free exact-two-component (X2C) Hamiltonian.\cite{Dyall:1997p9618,Kutzelnigg:2005p241102,Liu:2006p44102,Ilias:2007p64102,Cheng:2011p244112}
The experimental spectra were digitized using WebPlotDigitizer.\cite{WebPlotDigitizer}

%%%%%%%%%%%%%%%%%%%%%%%%%%%%%%%%%%%%%%%%%%%%%%%%%%%%%%%%%%%%%%%%%%%%%
%% 5. Results and Discussion
%% 5.1. XPS Spectra of substituted Ferrocenes
%% 5.2. XPS Spectra of the Azobenzene and its dynamics
%%%%%%%%%%%%%%%%%%%%%%%%%%%%%%%%%%%%%%%%%%%%%%%%%%%%%%%%%%%%%%%%%%%%%

\section{Results and Discussion}
\label{sec:results}

\subsection{Substituted Ferrocenes}
\label{sec:results:ferrocenes}

\begin{figure*}[t!]
	\centering
	\includegraphics[width=1.0\textwidth]{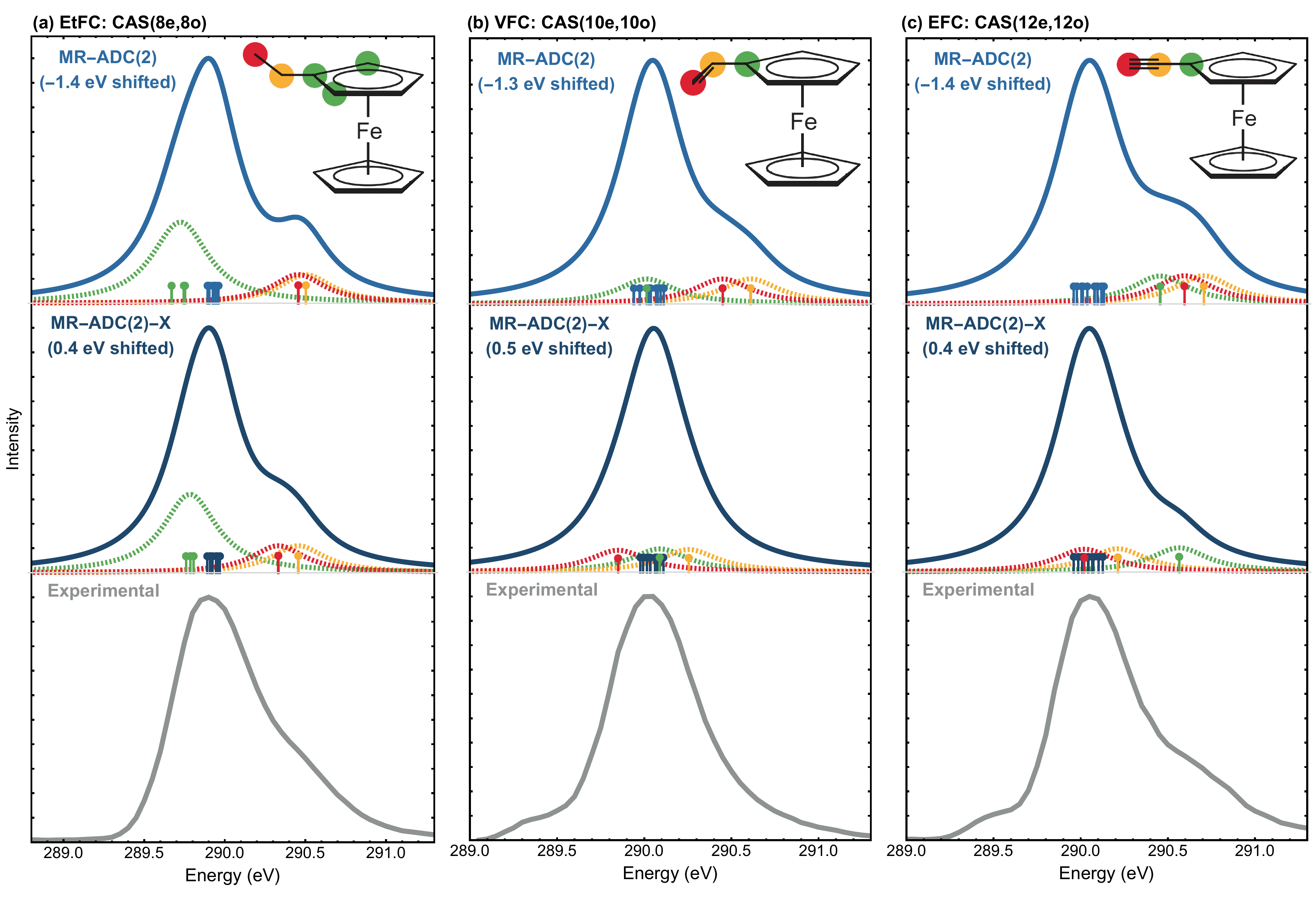}
	\caption{
		Carbon K-edge XPS spectra of (a) ethyl-ferrocene (EtFC), (b) vinyl-ferrocene (VFC), and (c) ethynyl-ferrocene (EFC) computed using the MR-ADC methods compared to the experimental results.\cite{Boccia:2012p134308}
		0.20 eV broadening parameter and were shifted to align with the peak maxima with the those in the experimental spectra.
		See \cref{sec:comp_details} for computational details.
		Experimental spectra were reprinted from Ref.\@ \citenum{Boccia:2012p134308}, with the permission of AIP Publishing.
		}
	\label{fig:Ferrocenes_XPS_plot}
\end{figure*}

We first apply the efficient MR-ADC implementation developed in this work to three ferrocene derivatives with the ethyl (Et), vinyl (V), or ethynyl (E) substituent (R) in one of the cyclopentadienyl (Cp) rings (\cref{fig:Ferrocenes_XPS_plot}).
The redox potentials of these molecules are highly sensitive to the R group, making them attractive surface modifiers for metals and semiconductors in memory devices, electrochemical sensors, batteries, and fuel cells.\cite{Roth:2003p505,Lindsey:2011p638,Heinze:2013p5623,Fabre:2016p4808,Rauf:2023p5765}
Meanwhile, the properties of ferrocene and its derivatives are known to be strongly influenced by electron correlation effects and have been the focus of many theoretical studies.\cite{Xu:2003p2716,Coriani:2006p245,Harding:2008p64,DeYonker:2009p24106,Phung:2012p883,Burow:2014p180,Wang:2016p4833,YanezS:2017p65,Sand:2017p34101,Sayfutyarova:2017p4063,Irfan:2020p1,Trivedi:2023p15929,Wang:2023p}
The carbon K-edge XPS spectra of EtFC, VFC, and EFC in the gas phase have been measured experimentally,\cite{Boccia:2012p134308} allowing us to test the performance of MR-ADC for these challenging systems.

\cref{fig:Ferrocenes_XPS_plot}a shows the experimental C K-edge XPS  spectrum of ethyl-ferrocene (EtFC)\cite{Boccia:2012p134308} along with the theoretical spectra calculated at the MR-ADC(2) and MR-ADC(2)-X levels of theory.
The experimental XPS spectrum shows a broad feature with a maximum at 290.0 eV and a shoulder around 290.5 eV.
The MR-ADC(2) and MR-ADC(2)-X calculations reproduce the experimental spectrum quite well when supplied with 0.20 eV broadening.
The best agreement with the experiment is shown by MR-ADC(2)-X, which underestimates the band maximum by 0.4 eV and accurately describes the relative energy of the shoulder feature.
MR-ADC(2) overestimates the band maximum by 1.4 eV and exhibits a somewhat more pronounced shoulder.

Analyzing the results of MR-ADC calculations reveals that the lowest-energy core-ionized states in the EtFC XPS spectrum are localized on the three C atoms of the substituted Cp ring directly adjacent to the ethyl (R = Et) group (\cref{fig:Ferrocenes_XPS_plot}a).
The energetic stabilization of these states is consistent with increased core-hole screening due to the donation of electron density from Et to Cp, as evidenced by the preferential localization of occupied CASSCF natural orbitals on the carbons nearest to Et (Figure S2 of Supporting Information). 
The core-ionized states localized on the remaining two C atoms of substituted Cp and all carbons of the unsubstituted Cp show very similar ionization energies, $\sim$ 0.2 eV higher than the lowest C K-edge ionization threshold.
Together, the Cp carbons give rise to a strong peak observed in the experimental XPS spectrum at 290.0 eV.
The shoulder feature at 290.5 eV originates from ionizing the ethyl group, which requires additional $\sim$ 0.5 eV of energy due to the depletion of  substituent electron density  and less efficient core-hole screening.

\cref{fig:Ferrocenes_XPS_plot}b presents the experimental\cite{Boccia:2012p134308} and simulated C K-edge XPS spectra of vinyl-ferrocene (VFC).
In contrast to EtFC, no shoulder feature with significant intensity is observed at $\sim$ 290.5 eV in the experimental spectrum, indicating smaller variations in the electronic density of C atoms.
A weak shoulder-like feature is seen at $\sim$ 289.5 eV.
MR-ADC(2)-X predicts a single spectral band with similar ionization energies for most C atoms.
The occupied CASSCF natural orbitals of VFC (Figure S3) exhibit a $\pi$-delocalization of electron density between the vinyl group and substituted Cp ring, resulting in a more even core-hole screening among the C K-edge ionized states. 
The largest difference in the K-edge ionization energy is observed between the \ch{CH2} and \ch{CH} groups of the vinyl substituent (\cref{fig:Ferrocenes_XPS_plot}b), which give rise to the lowest- and highest-energy transitions in the MR-ADC(2)-X spectrum, respectively.
The significant difference in carbon core-hole screening of these groups can be attributed to the resonance 
\begin{center}
	\includegraphics[width=0.4\textwidth]{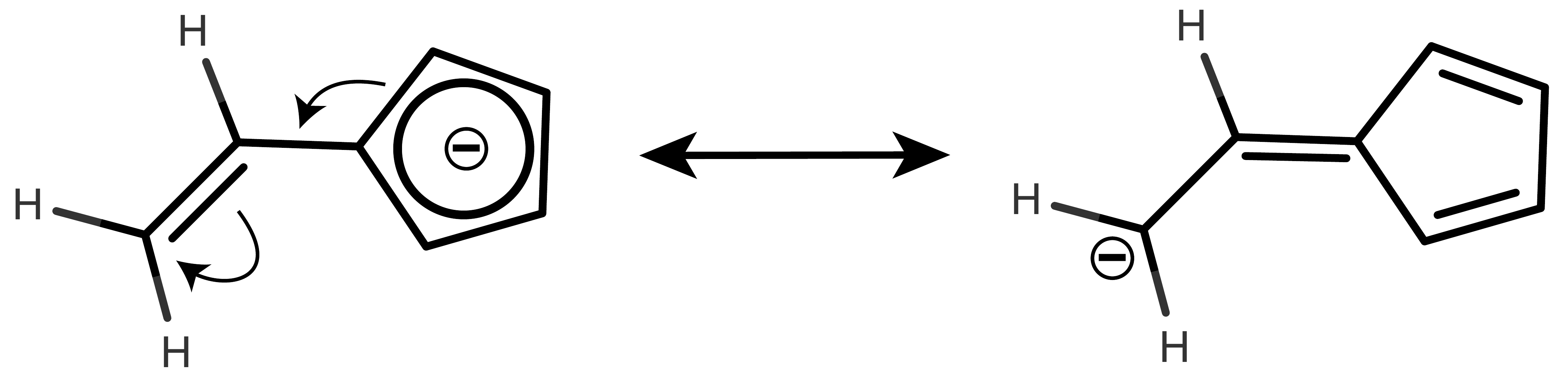}
\end{center}
and is consistent with the enhanced $\pi$-delocalization of CASSCF natural orbitals (Figure S3). 
In particular, the ionization of \ch{CH2} group in the vinyl substituent may be responsible for the appearance of weak shoulder at $\sim$ 289.5 eV in the experimental spectrum,\cite{Boccia:2012p134308} although no clear shoulder is seen in the MR-ADC(2)-X spectrum.
MR-ADC(2) overestimates the peak position and spacing relative to the experiment, predicting a weak shoulder feature with the selected broadening parameter.
This discrepancy can be attributed to the overestimation of core ionization energy for the terminal carbon atom of the vinyl group and is likely associated with the less accurate description of excited-state orbital relaxation effects in MR-ADC(2) compared to MR-ADC(2)-X.

Finally, we consider the experimental\cite{Boccia:2012p134308} and simulated C K-edge XPS spectra of ethynyl-ferrocene (EFC) shown in \cref{fig:Ferrocenes_XPS_plot}c.
Similar to EtFC, the EFC core ionization spectra exhibit a strong peak followed by a shoulder feature $\sim$ 0.5 eV higher in energy.
In addition, a weak shoulder is again observed at $\sim$ 289.5 eV, similar to the one in the VFC spectrum. 
As for EtFC and VFC, MR-ADC(2)-X has the best agreement with the experiment, slightly underestimating the intensity of shoulder at $\sim$ 290.5 eV.
In contrast to EtFC where the $\sim$ 290.5 eV shoulder was assigned to the C atom of substituent, the MR-ADC(2)-X calculations suggest that this feature in the EFC spectrum originates from the Cp carbon directly bonded to the ethynyl group. 
These results suggest that ethynyl acts as an acceptor, depleting the electron density on the Cp carbon adjacent to the substituent and weakening the screening of corresponding core-ionized state. 
Indeed, the analysis of CASSCF natural orbitals (Figure S4) shows a significant occupation of in-plane and out-of-plane $\pi$-antibonding ethynyl orbitals, indicating noticeable back-donation from Cp to the ethynyl group.
MR-ADC(2) shows a larger error in the peak maximum and predicts a more intense shoulder than what is observed in the experiment as a result of overestimating the relative energy of core-hole states in the substituent.
As for VFC, the MR-ADC methods do not predict a weak shoulder at $\sim$ 289.5 eV, which was assigned to the terminal carbon of ethynyl group in the experimental study.\cite{Boccia:2012p134308}
Although this error may be associated with insufficient description of orbital relaxation effects, it may also originate from the lack of vibrational effects in the simulation and requires further study.

Overall, our results demonstrate that the efficient MR-ADC implementation developed in this work can routinely simulate the K-edge XPS spectra of transition metal complexes with large basis sets.
Specifically, the MR-ADC(2) and MR-ADC(2)-X calculations reported here were performed correlating all electrons in 1412 (EFC), 1472 (VFC), and 1532 (EtFC) molecular orbitals.
Using a single Intel Xeon Gold 6148 computer node with 40 CPUs, the wall time of MR-ADC(2) and MR-ADC(2)-X simulations did not exceed 4 and 40 hours, respectively, after completing the reference CASSCF calculations. 
Consistent with our earlier benchmarks,\cite{Moura:2022p4769,Moura:2022p8041} the MR-ADC(2)-X results are in very good agreement with the experimental gas-phase XPS spectra.

\subsection{Azobenzene Photoisomerization}
\label{sec:results:azobenzene}

As a multireference approach, MR-ADC is well suited for predicting and interpreting the transient spectra in TR-XPS measurements where X-ray or extreme ultraviolet light is used to probe the electronic structure and molecular dynamics in an excited state populated with a UV/Vis pump photon.
We have recently demonstrated this capability by simulating the transient XPS spectra of \ch{Fe(CO)5} and its photodissociation products (\ch{Fe(CO)4}, \ch{Fe(CO)3}) following the excitation with 266 nm pump where MR-ADC provided insights into the origin of chemical shifts observed in the experiment.\cite{Gaba:2024p15927}

Here, we use our efficient implementation of MR-ADC to simulate the ground- and excited-state XPS signatures along the photoisomerization of azobenzene (\ch{(C6H5)2N2}).
Azobenzene is a photoswitch molecule that converts from the lowest-energy \textit{trans}- to the higher-energy \textit{cis}-isomer upon irradiation with ultraviolet ($\sim$ 365 nm) light.\cite{Schultz:2003p8098,Crespi:2019p133,Jerca:2021p51}
The precise mechanism of \textit{trans}-\textit{cis} photoisomerization has been a matter of debate,\cite{Rau:1982p1616,Lednev:1998p9161,Lednev:1998p68,Schultz:2003p8098,Conti:2008p5216,Quick:2014p8756,Tavadze:2018p285} with some studies suggesting that it involves initial excitation to the $\pi\pi^*$ excited state, followed by rapid internal conversion to the $n\pi^*$ potential energy surface where the isomerization can take place.\cite{Cembran:2004p3234,Bandara:2011p1809,Tan:2015p5860,Casellas:2016p3068,Nenov:2018p1534,Aleotti:2019p6813,Yu:2020p20680,Merritt:2021p19155}
Time-resolved photoelectron spectroscopy (TR-PES) measurements in the UV region of electromagnetic spectrum provided valuable insights about the mechanism of this photoisomerization.\cite{Schultz:2003p8098}
Complimentary to TR-PES, TR-XPS can help to elucidate this mechanism further by detecting the element-specific transient spectral signatures along the course of photochemical reaction.
Although no experimental TR-XPS studies have been presented to date, the ground-state carbon and nitrogen K-edge XPS spectra of {\it trans}-azobenzene in the gas phase have been recently reported.\cite{Carlini:2023p54201}
In addition to TR-XPS, time-resolved X-ray absorption spectroscopy has been proposed and theoretically evaluated as a technique for studying azobenzene photoisomerization.\cite{Ehlert:2018p144112,Segatta:2019p245}

\begin{figure*}[t!]
	\centering
	\includegraphics[width=1.0\textwidth]{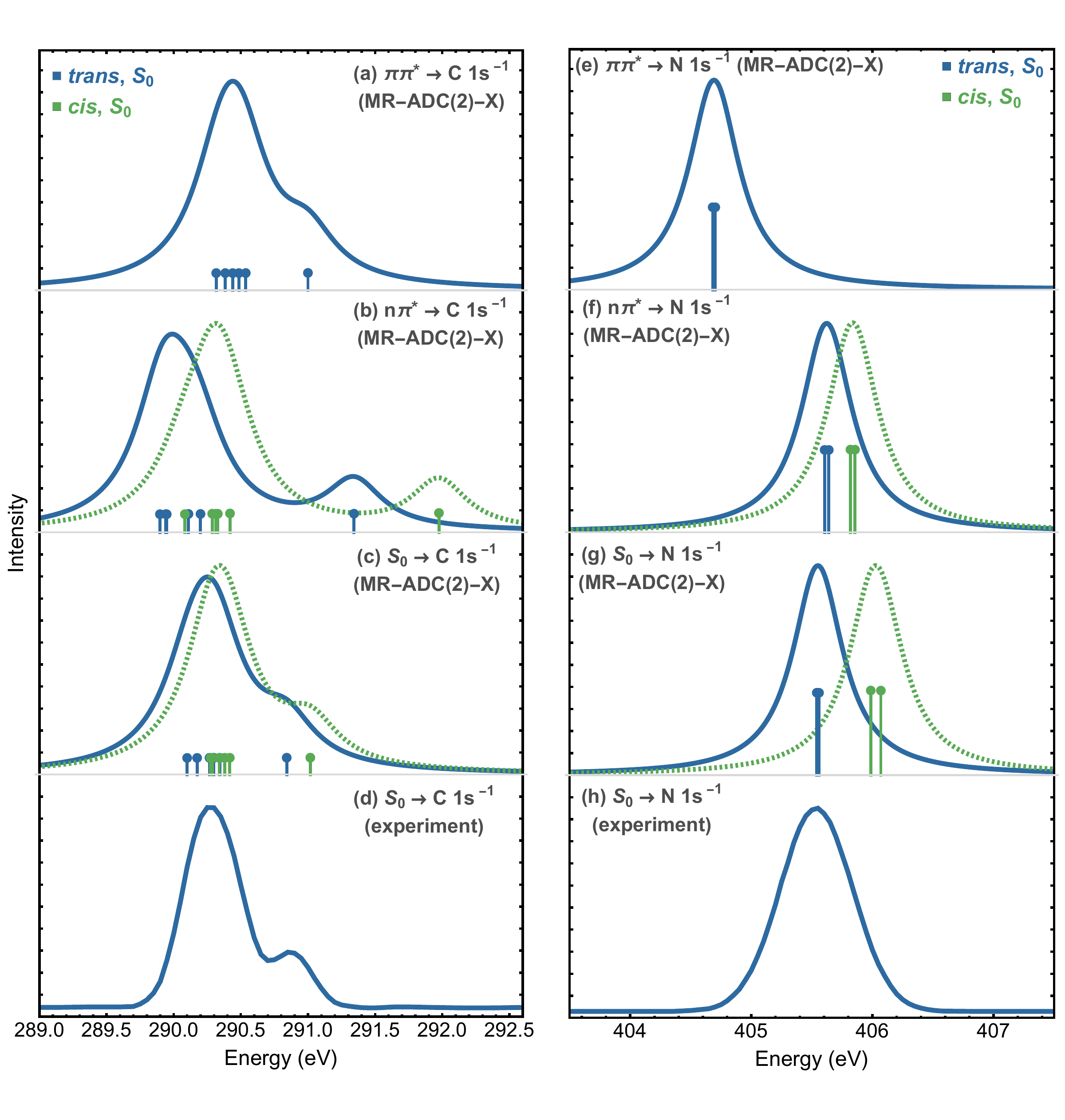}
	\caption{
		Carbon and Nitrogen K-edge XPS spectra of azobenzene simulated for the $\pi\pi^\ast$ excited (a,e), $n\pi^\ast$ excited (b,f), and $S_0$ ground (c,g) states using the MR-ADC(2)-X method.
		The results are calculated at the {\it trans} (blue) or {\it cis} (green) ground-state equilibrium geometries.
		The simulated spectra were broadened with a 0.25 eV parameter and shifted by 0.85 eV (a--c) or 0.63 eV (e--g).
		See \cref{sec:comp_details} for computational details.
		Also shown are the experimental $S_0$ XPS spectra of {\it trans-}azobenzene (d,h), reprinted from Ref.\@ \citenum{Carlini:2023p54201}, with the permission of AIP Publishing.
	}
	\label{fig:Azo_XPS_MRADC2X_plot}
\end{figure*}

We performed the MR-ADC(2)-X calculations of XPS spectra at three azobenzene geometries: the equilibrium structures of \textit{trans}- and \textit{cis}-isomers and the geometry of minimum energy conical intersection (MECI) between the two lowest-energy singlet states that is believed to be important in photoisomerization (\cref{sec:comp_details}).\cite{WeiGuangDiau:2004p950,Conti:2008p5216,Yu:2020p20680}
For each structure, the carbon and nitrogen K-edge XPS signatures were simulated for the ground ($S_0$) and lowest-energy excited ($S_1$) singlet states.
For the \textit{trans}- and \textit{cis}-isomers, the $S_1$ state corresponds to the $n\pi^\ast$ excitation from the molecular orbital localized on nitrogen lone pairs to the lowest-energy $\pi$-antibonding orbital of the molecule (Figures 4b, S6, S9).
We also computed the XPS spectra of \textit{trans}-azobenzene in the second singlet excited state ($S_2$) corresponding to the $\pi\pi^\ast$  electronic transition (Figures 4c and S7).
The calculations employed large quadruple-zeta basis sets correlating all electrons in 1476 molecular orbitals and CAS(16e,15o).
Additional computational details can be found in \cref{sec:comp_details} and Supplementary Information.
We note that our calculations did not incorporate excited-state relaxation  and nuclear dynamics effects that can be important for accurate interpretation of TR-XPS spectra.

\begin{figure*}[t!]
	\centering
	\includegraphics[width=1.0\textwidth]{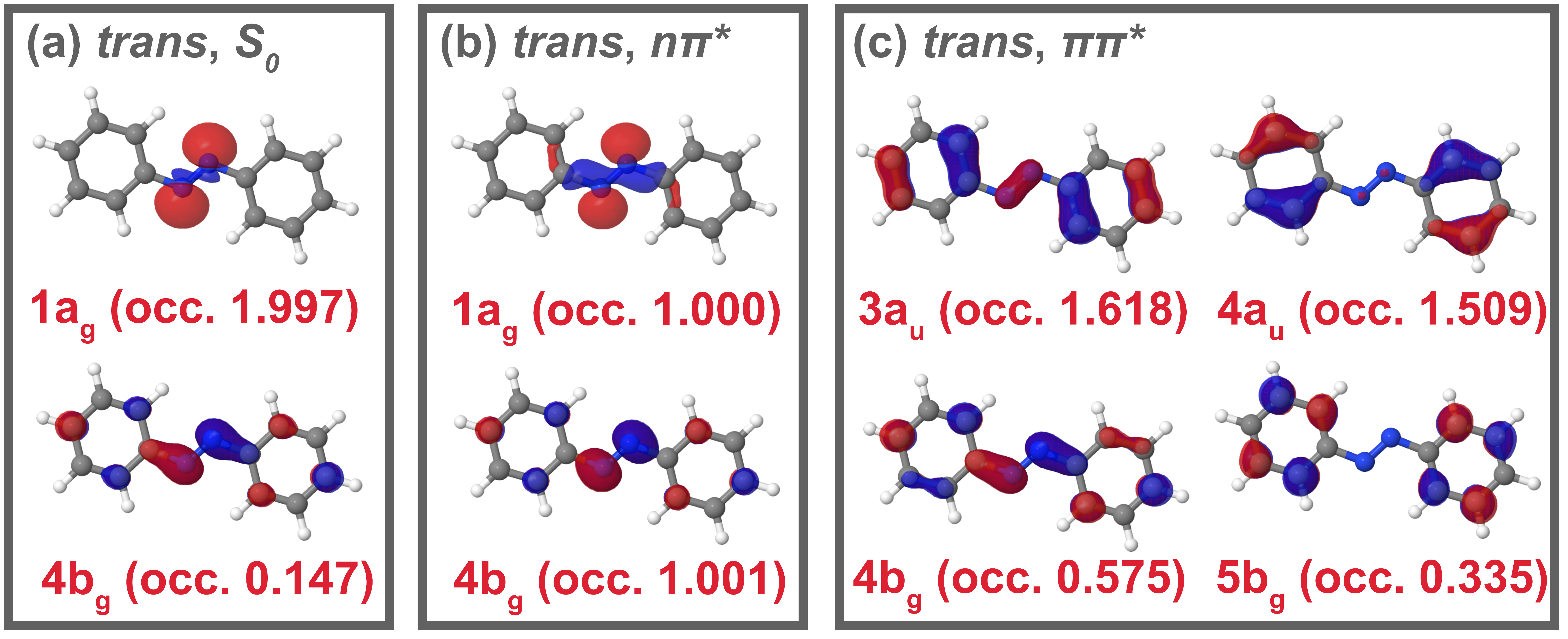}
	\caption{
	Selected CASSCF natural orbitals and their occupations for the three electronic states of \textit{trans}-azobenzene: (a) ground state ($S_0$), (b) $n\pi^\ast$ excited state, and (c) $\pi\pi^\ast$ excited state.
	Calculations were performed using the (16e,15o) active space and the cc-pwCVQZ basis set.
	See Supplementary Information for the natural orbital plots of remaining active orbitals.
	}
	\label{fig:Azo_NATORB_CAS1615}
\end{figure*}

Figures 3a to 3c show the simulated C K-edge XPS spectra of \textit{trans}- and \textit{cis}-azobenzene in the $S_0$, $n\pi^\ast$, and $\pi\pi^\ast$  electronic states.
For the ground state of \textit{trans}-isomer, the results of simulations are in an excellent agreement with the experimental spectrum (Figure 3d) measured by Carlini et al.\cite{Carlini:2023p54201}
In all C K-edge XPS spectra, an intense peak is followed by a feature with weaker intensity that appears as a shoulder in the $S_0$ spectrum.
These two features correspond to the K-edge ionization of C atoms in two distinct chemical environments: 1) ten H-bonded C's and 2) two N-bonded C's.
Due to increasing electronegativity ($\chi$) in the order $\chi$(H) $<$ $\chi$(C) $<$ $\chi$(N), the N-bonded carbons exhibit a higher charge and less efficient core-hole screening than those attached to the H atoms, resulting in a significant blueshift of the corresponding feature.
As can be seen in \cref{fig:Azo_XPS_MRADC2X_plot}c, the ground-state XPS spectra of \textit{trans}- and \textit{cis}-isomers exhibit nearly identical spacing between the two peaks ($\sim$ 0.6 eV), indicating that the electronic environment of C atoms in these states is similar.

Analyzing the $n\pi^\ast$ and $\pi\pi^\ast$ C K-edge XPS spectra reveals significant differences in the electronic structure of these excited states.
The $\pi\pi^\ast$ spectrum is very similar to that of the $S_0$ state, suggesting that the $S_0 \rightarrow \pi\pi^\ast$ transition does not significantly change the electron density distribution around the C atoms of \textit{trans}-azobenzene.
In contrast, populating the $n\pi^\ast$ state results in $\sim$ 0.3 eV redshift of the lowest-energy peak and $\sim$ 0.4 eV blueshift of the peak with a weaker intensity, increasing their spacing from $\sim$ 0.6 to $\sim$ 1.3 eV in both \textit{trans}- and \textit{cis}-azobenzene.
These results are consistent with the analysis of CASSCF natural orbitals (\cref{fig:Azo_NATORB_CAS1615}) where the $n\pi^\ast$ excited state shows a higher delocalization of the $\pi^\ast$ ($4b_g$) orbital as compared to that of the $n$ ($1a_g$) orbital, transferring some electron density into the phenyl rings of azobenzene.
On the contrary, the orbitals involved in the $\pi\pi^\ast$ excitation are similarly delocalized across the C framework of the molecule.
Importantly, the results of calculations demonstrate that the $\pi\pi^\ast$ and $n\pi^\ast$ states exhibit noticeably different C K-edge XPS spectra at the \textit{trans}-geometry, which may facilitate their spectroscopic identification in the TR-XPS experiments.

Figures 3e to 3g present the N K-edge XPS spectra of \textit{trans}- and \textit{cis}-azobenzene.
For each structure and electronic state, the N K-edge spectrum displays a single peak corresponding to the 1s ionization of two symmetry-equivalent nitrogen atoms.
The MR-ADC(2)-X method underestimates the experimental core binding energy\cite{Carlini:2023p54201} of \textit{trans}-azobenzene in the $S_0$ state by $\sim$ 0.6 eV.
In contrast to the C K-edge XPS spectra, the computed N K-ionization energies are more sensitive to the \textit{cis}/\textit{trans}-orientation of ground state geometry or excitation to the $\pi\pi^\ast$  state.
Specifically, isomerization from \textit{trans}- to \textit{cis}-azobenzene on the  $S_0$ potential energy surface increases the N K-binding energy by $\sim$ 0.5 eV while the $\pi\pi^\ast$ excitation at the \textit{trans} geometry decreases it by $\sim$ 0.8 eV.
Exciting the molecule to the $n\pi^\ast$ state results in $\sim$ 0.2 and 0.4 eV blueshifts in peak position for the \textit{trans}- and \textit{cis}-isomers, respectively.
As for the C K-edge, all spectral changes can be interpreted based on the analysis of CASSCF natural orbitals (\cref{fig:Azo_XPS_MRADC2X_plot}).
For example, the significant ($\sim$ 0.8 eV) blueshift of N K-signal for the $\pi\pi^\ast$ state is consistent with excited-state electron density concentrating on the orbitals localized on the N atoms ($3a_u$ and $4b_g$, \cref{fig:Azo_XPS_MRADC2X_plot}c) as opposed to the orbitals localized on the phenyl rings ($4a_u$ and $5b_g$).

\begin{figure*}[t!]
	\centering
	\includegraphics[width=1.0\textwidth]{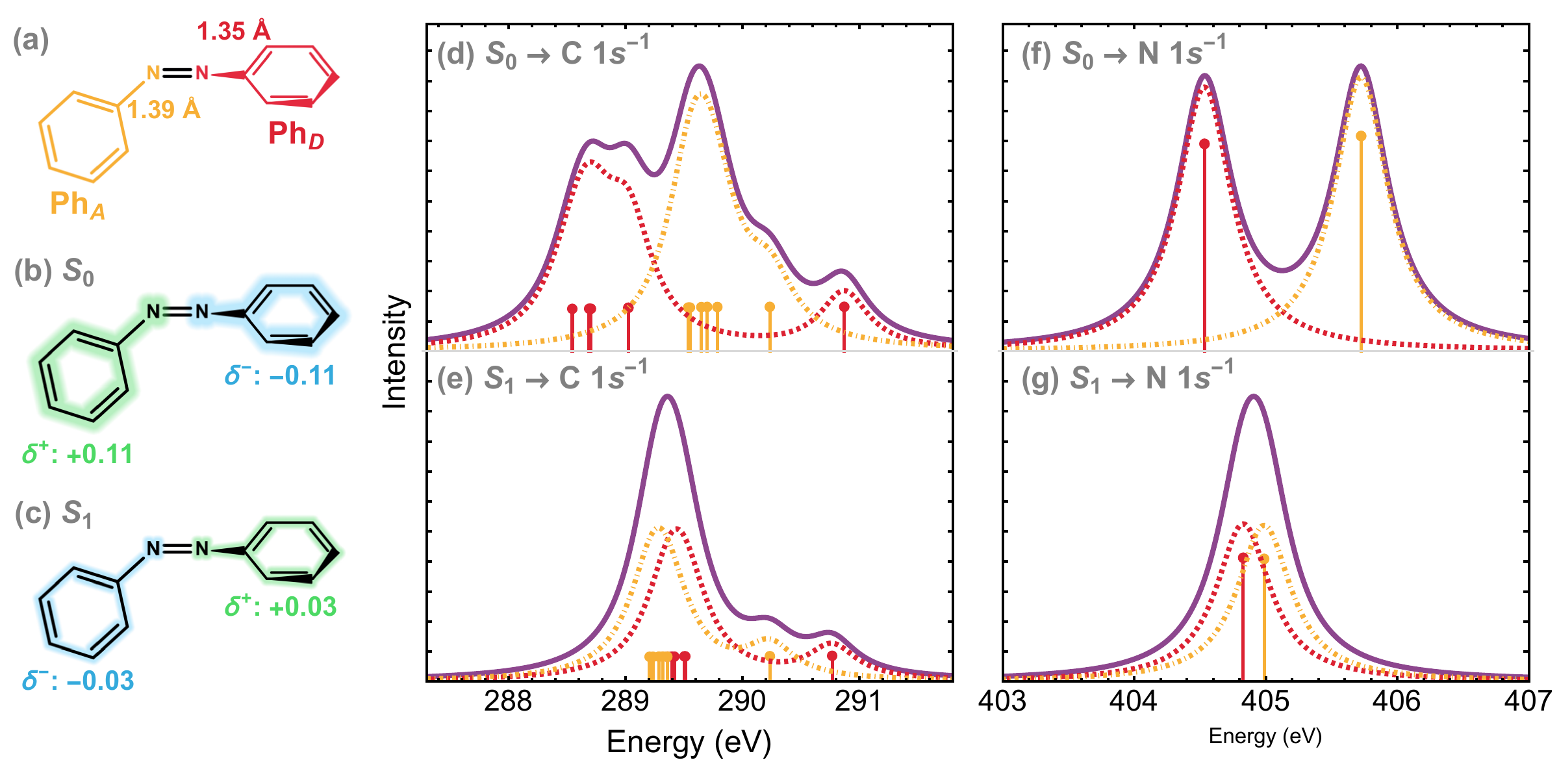}
	\caption{
		(a) Schematic representation of the azobenzene molecule at the $S_0$--$S_1$ minimum energy conical intersection geometry.
		(b) and (c) Mulliken charges of the NPh$_A$ and NPh$_D$ fragments computed for the $S_0$ and $S_1$ states, respectively.
		(d) and (e) C K-edge XPS spectra for the $S_0$ and $S_1$ states, respectively.
		(f) and (g) N K-edge XPS spectra for the $S_0$ and $S_1$ states, respectively.
		All calculations used the SA2-CASSCF(16e,15o) reference wavefunction and the cc-pwCVQZ basis set with X2C relativistic corrections.
		The spectra were simulated using MR-ADC(2)-X and 0.25 eV broadening parameter.
	}
	\label{fig:Azo_CI_XPS_plot}
\end{figure*}

Finally, to demonstrate an application of MR-ADC at non-equilibrium regions of potential energy surfaces, we analyze the XPS spectra computed for the $S_0$--$S_1$ MECI geometry shown in \cref{fig:Azo_CI_XPS_plot}a.
The MECI exhibits an unusual structure of $C_1$ symmetry with a 91.6$^{\degree}$ dihedral angle between the two phenyl rings (Ph) and significantly different N--C bond distances (1.39 and 1.35 \AA).
As can be seen from the occupations of CASSCF natural orbitals (Figures S10 and S11), the two singlet states participating in MECI show significantly different degree of open-shell character.
Consistent with Ref.\@ \citenum{Segatta:2019p245}, we will refer to the state with lower/higher open-shell character as ``closed-shell''/``open-shell'' and label them as $S_0$/$S_1$.
To distinguish between  the two non-equivalent NPh fragments, we denote the Ph group with the shorter/longer N--C bond as  Ph$_{D}$/Ph$_{A}$.
The Mulliken analysis of charge density (\cref{fig:Azo_CI_XPS_plot}b,c) reveals that NPh$_{D}$ has a lower total charge than NPh$_{A}$ in the $S_0$ state, but a higher charge in the $S_1$ state.

The C and N K-edge XPS spectra of $S_0$ and $S_1$ simulated at the MECI geometry are shown in Figures 5d to 5g.
Due to the asymmetric structure, the spectra show distinct XPS signatures for the C and N atoms of the NPh$_{D}$ and NPh$_{A}$ groups of the molecule.
In the closed-shell $S_0$ state, the lowest-energy C and N K-edge ionization occurs in the Ph$_{D}$ fragment, which exhibits significantly higher electron density and more efficient core-hole screening.
Ionizing the Ph$_{A}$ group requires $\sim$ 1 eV of additional ionization energy, significantly broadening the C and N K-edge XPS spectra in comparison to those of ground-state azobenzene (\cref{fig:Azo_XPS_MRADC2X_plot}).
The XPS signatures of open-shell $S_1$ state more closely resemble the ground-state spectra with overlapping peaks originating from Ph$_{D}$ and Ph$_{A}$.
These results are consistent with the analysis of Mulliken charges (Figures 5b and 5c) and CASSCF natural orbitals (Figures S10 and S11), which reveal a more even electron density distribution between NPh$_{D}$ and NPh$_{A}$ in the $S_1$ state.

Overall, the results of our simulations suggest that TR-XPS can be a useful tool in investigating the photoisomerization of azobenzene molecule, with complementary information provided by the measurements at C and N K-edges.
In particular, the C K-edge spectra are expected to help with the detection of molecules in the $n\pi^*$ excited state and are rather insensitive to their \textit{trans}/\textit{cis}-orientation.
On the contrary, the peak shifts measured in the N K-edge spectra can be useful to identify molecules in the $\pi\pi^*$ state and to distinguish between the spectral signatures of \textit{trans}- and \textit{cis}-isomers.
Monitoring the broadening of C and N K-edge XPS spectra as a function of time may help to provide additional details about the mechanism of photoisomerization.
In addition to the XPS spectra for each electronic state (\cref{fig:Azo_XPS_MRADC2X_plot,fig:Azo_CI_XPS_plot}), we computed the difference spectra (Figures S13 to S16) that may be helpful in interpreting the TR-XPS measurements.
We stress, however, that our simulations did not incorporate the excited-state relaxation and nuclear dynamics effects, which can be important for the accurate interpretation of TR-XPS spectra and will be the subject of future work.

\section{Conclusions}
\label{sec:conclusions}

In this work, we presented an efficient implementation of multireference algebraic diagrammatic construction theory with core-valence separation for simulating core-ionized states and X-ray photoelectron spectra (CVS-IP-MR-ADC).
Developed in the open-source and freely available Prism program,\cite{Prism} CVS-IP-MR-ADC takes advantage of spin adaptation, automatic code generation, and efficient handing of two-electron integrals via density fitting.
Incorporating dynamic and static correlation, CVS-IP-MR-ADC allows to accurately simulate the X-ray or extreme ultraviolet photoelectron spectra (XPS) for molecules in excited electronic states or at nonequilibrium ground-state geometries. 
The CVS-IP-MR-ADC is resilient to intruder-state problems and allows to calculate many (10's or even 100's) excited states starting with a single CASSCF wavefunction for the reference electronic state.

We demonstrated the capabilities of our efficient CVS-IP-MR-ADC implementation by applying it to substituted ferrocene complexes and azobenzene molecule along its photoisomerization pathway. 
In all calculations, we used core-polarized quadruple-zeta basis sets, correlating all electrons in more than 1500 molecular orbitals. 
For the ground electronic states of substituted ferrocenes and {\it trans-}azobenzene, the carbon K-edge XPS spectra simulated using the extended second-order CVS-IP-MR-ADC method (CVS-IP-MR-ADC(2)-X) are in a very good agreement with experimental measurements. 
Encouraged by these results, we also computed the carbon and nitrogen K-edge XPS spectra for the azobenzene molecule in its excited states and at the geometry of asymmetric minimum energy conical intersection, which may be useful in interpreting the future time-resolved XPS experiments.

The efficient implementation strategy presented in this work can be used to develop fast MR-ADC methods for simulating other spectroscopic properties, including electron attachment and ionization in the UV/Vis region,\cite{Chatterjee:2019p5908,Chatterjee:2020p6343} and neutral excitations in the UV/Vis and X-ray absorption spectroscopies (XAS).\cite{Sokolov:2018p204113,Mazin:2021p6152,Mazin:2023p4991} 
In addition, spin--orbit coupling effects can be incorporated to enable accurate simulations of L- and M-edge XPS and XAS spectra.\cite{Gaba:2024p15927}
These developments will be reported in the forthcoming publications of this series and will further expand the application domain of MR-ADC.

\begin{acknowledgement}
This work was supported by the National Science Foundation, under Grant No. CHE-2044648.
The authors would like to thank Nicholas Gaba for testing the CVS-IP-MR-ADC code and valuable suggestions on its improvements.
Computations were performed at the Ohio Supercomputer Center under Project No.\@ PAS1583.\cite{OhioSupercomputerCenter:1987}
\end{acknowledgement}

\begin{suppinfo}
Relationships between spin-orbital and spin-free 3- and 4-RDMs, benchmark of density fitting approximation in the CVS-IP-MR-ADC calculations, geometries of all molecules, CASSCF natural orbitals, tables with XPS ionization energies and spectroscopic factors, and simulated difference XPS spectra for azobenzene. 
\end{suppinfo}

\providecommand{\latin}[1]{#1}
\makeatletter
\providecommand{\doi}
  {\begingroup\let\do\@makeother\dospecials
  \catcode`\{=1 \catcode`\}=2 \doi@aux}
\providecommand{\doi@aux}[1]{\endgroup\texttt{#1}}
\makeatother
\providecommand*\mcitethebibliography{\thebibliography}
\csname @ifundefined\endcsname{endmcitethebibliography}
  {\let\endmcitethebibliography\endthebibliography}{}

\end{document}